\documentclass{article}
\usepackage{spconf,amsmath,graphicx}

\usepackage{multirow}
\usepackage{graphicx}
\usepackage{amssymb, amsmath, bm, mathtools,array}
\usepackage{textcomp}
\usepackage{hyperref}
\usepackage{subfig}
\usepackage{verbatim,lipsum}
\usepackage{multicol}
\RequirePackage[normalem]{ulem} 
\RequirePackage{color}\definecolor{BLUE}{rgb}{0,0.20,0.75} 
\RequirePackage{color}\definecolor{BROWN}{RGB}{60,128,49} 

\newcommand{\bs}[1]{\boldsymbol{#1}}
\title{A comparison of recent waveform generation and acoustic modeling methods for neural-network-based speech synthesis}
\name{Xin Wang$^1$, Jaime Lorenzo-Trueba$^1$, Shinji Takaki$^1$, Lauri Juvela$^2$, Junichi Yamagishi$^{1}$\sthanks{This work was partially supported by MEXT KAKENHI Grant Numbers (15H01686, 16H06302, 17H04687), and JST ACT-I Grant Number (JPMJPR16UG).}}
\address{$^1$National Institute of Informatics, Japan \\
  $^2$Aalto University, Finland \\
  {\small \tt wangxin@nii.ac.jp, jaime@nii.ac.jp, takaki@nii.ac.jp, lauri.juvela@aalto.fi,  jyamagis@nii.ac.jp}}

\begin{document}
\ninept

\onecolumn
{\noindent\Large \textbf{IEEE Copyright Notice}}

${}$

{\noindent\large \copyright 2018 IEEE. 
Personal use of this material is permitted. Permission from IEEE must be obtained for all other uses, in any current or future media, including reprinting/republishing this material for advertising or promotional purposes, creating new collective works, for resale or redistribution to servers or lists, or reuse of any copyrighted component of this work in other works.

${}$

\noindent
Accepted to be published in: Proceedings of the 2018 IEEE International Conference on Acoustics, Speech, and Signal Processing, April 16-20, 2018, Calgary, Canada}
\twocolumn
\maketitle
\begin{abstract}
Recent advances in speech synthesis suggest that limitations such as the lossy nature of the amplitude spectrum with minimum phase approximation and the over-smoothing effect in acoustic modeling can be overcome by using advanced machine learning approaches. In this paper, we build a framework in which we can fairly compare new vocoding and acoustic modeling techniques with conventional approaches by means of a large scale crowdsourced evaluation. Results on acoustic models showed that generative adversarial networks and an autoregressive (AR) model performed better than a normal recurrent network and the AR model performed best. Evaluation on vocoders by using the same AR acoustic model demonstrated that a Wavenet vocoder outperformed classical source-filter-based vocoders. 
Particularly, generated speech waveforms from the combination of AR acoustic model and Wavenet vocoder achieved a similar score of speech quality to vocoded speech.

\end{abstract}

\section{Introduction}
\label{sec:intro}

{
Text-to-speech (TTS) synthesis aims at converting a text string into a speech waveform. A conventional TTS pipeline normally includes a frond-end text-analyzer and a back-end speech synthesizer, each of which may include multiple modules for specific tasks. 
For example, the back-end based on statistical parametric speech synthesis (SPSS) uses statistical acoustic models to map linguistic features from the text-analyzer into compact acoustic features. It then uses a vocoder to generate a waveform from the acoustic features.

Recent TTS systems have well adopted to the impact of deep learning. 
First, frameworks different from the established TTS pipeline are defined.
One framework merges the text-analyzer and acoustic models into a single neural network (NN), where the input text is directly mapped into acoustic features,  e.g., by Tactron \cite{Wang2017} and Char2wav \cite{Sotelo2017Char2wavES}. Another framework is the unified back-end represented by Wavenet \cite{oord2016wavenet} that directly converts the linguistic features into a waveform. There are also novel NN back-ends without the use of normal acoustic features, vocoders \cite{Espic2017,Takaki2017}, or duration model \cite{Ronanki2017}.
In parallel to these new frameworks, modules for the conventional TTS pipeline have also been developed. For instance, acoustic models based on the generative adversarial network (GAN) \cite{kaneko2017generative} and autoregressive (AR) NN \cite{wangARRMDN} have been reported to alleviate the over-smoothing effect in acoustic modeling. 
There is also the Wavenet-based vocoder \cite{Tamamori2017} that uses deep learning rather than signal processing methods to attain waveform generation.

Given the recent progress with TTS, we think it is important to understand the pros and cons of the new methods on the basis of common conditions. 
As an initial step, we compared the new methods that could easily be plugged into the SPSS-based back-end of the classical TTS pipeline. 
These methods were divided into two groups, and thus the comparison included two parts.
The first part compared acoustic models based on four types of NN, including the normal recurrent neural network (RNN), shallow AR-based RNN (SAR), 
and two GAN-based postfilter networks \cite{kaneko2017generative} that were attached to the RNN and SAR. The second part compared waveforms generation techniques, including the high-quality source-filter vocoder WORLD \cite{morise2016world}, another synthesizer based on the log-domain pulse model (PML) \cite{degottex2017log}, WORLD plus the Griffin-Lim algorithm for phase recovery \cite{griffin1984signal}, and the Wavenet-based vocoder. 

Section~\ref{sec:2} explains the details on the selected methods used for comparison. Section~\ref{sec:exp} explains the experiments and Section~\ref{sec:con} draws conclusions. 
}

\begin{figure}[t]
\includegraphics[width=\columnwidth]{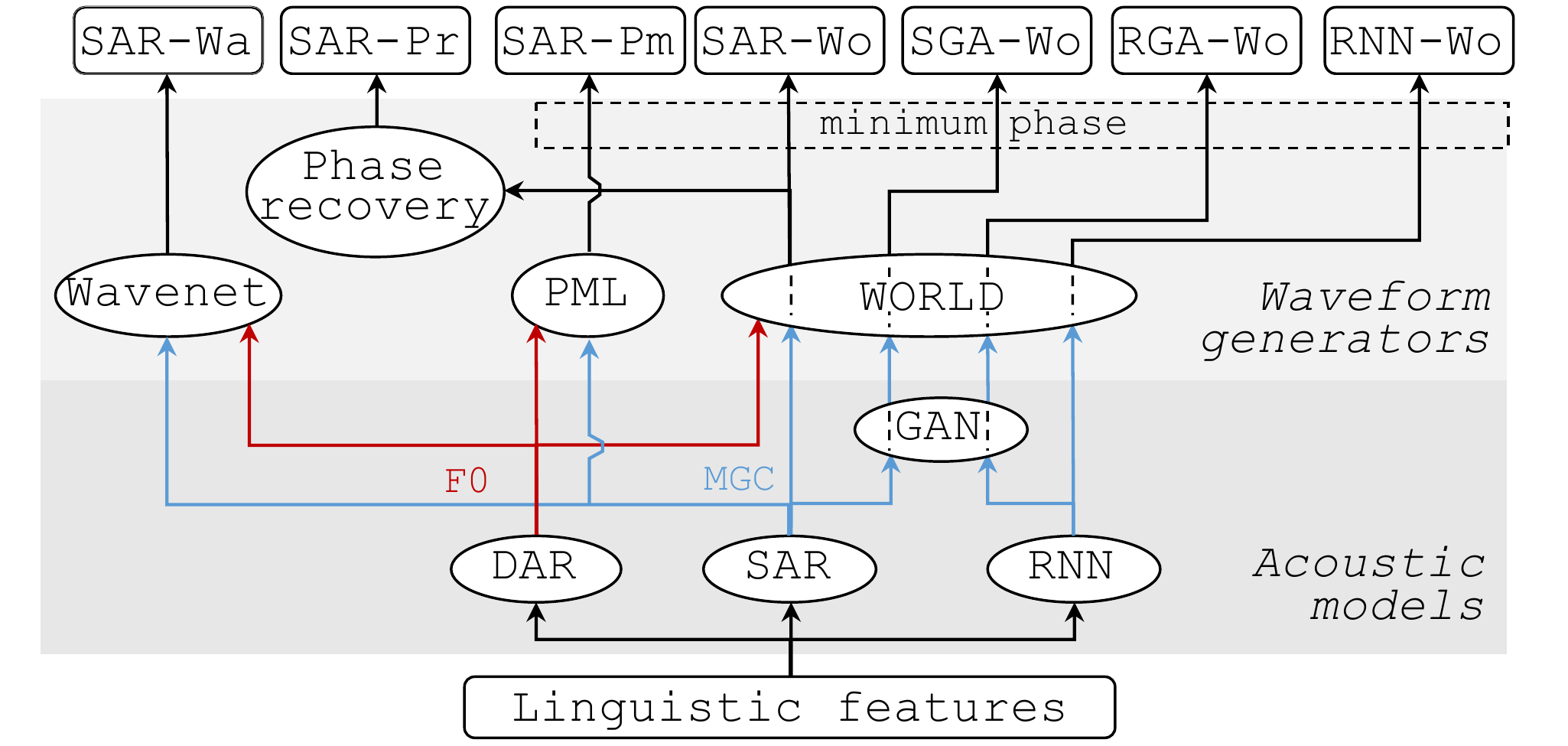}
\vspace{-7mm}
\caption{Flowchart for seven speech synthesis methods compared in this study. This figure hides the band aperiodicity (BAP) features and RNN that generates noise mask for {PML} vocoder.}
\vspace{-5mm}
\label{fig:systems}
\end{figure}

\vspace{-2mm}
\section{Speech synthesis methods}
\label{sec:2}
{
This comparison study was based on perceptual evaluation of synthetic speech. The selected acoustic models and waveform generation methods were combined into the seven speech synthesis methods given in Fig.~\ref{fig:systems}, where
 \texttt{SAR-Wa}, \texttt{SAR-Pm}, \texttt{SAR-Pr}, and \texttt{SAR-Wo} use the same acoustic models but different waveform generation modules, while \texttt{SAR-Wo}, \texttt{GAS-Wo}, \texttt{GAR-Wo}, and \texttt{RNN-Wo} use the same WORLD vocoder but differ in the acoustic model to generate the Mel-generalized coefficients (MGC) and band aperiodicity (BAP). The F0 for all the methods is generated by a deep AR (DAR) model \cite{wangDARF0}. The oracle duration is used for all the methods. Neither formant enhancement nor MLPG \cite{ref:tokuda00} was used.  

}

\subsection{Waveform generation}
\subsubsection{Relationship between acoustic features and waveforms}
\label{cep-conversion}
{
A vocoder in a conventional speech synthesis method synthesizes the waveform given the generated acoustic features by the acoustic model. However, the synthesized waveform may be unnatural even if the acoustic model is perfect because the acoustic features may be imperfectly defined.
For example, to extract cepstral features from the waveform, the first step is to apply discrete Fourier transform (DFT) to the windowed waveform and then acquire the complex spectrum, $\bs{s}_n = [s_{n,1}, \cdots, s_{n,F}] \in\mathbb{C}^F$.
Here, $s_{n,f}=a_{n,f} + ib_{n,f} \in\mathbb{C}$ is a complex number, $n$ is the index of a speech frame, and $f\in[1,F]$ is the index of a frequency bin. The next step is to compute the power spectrum, $\bs{p}_n=[p_{n,1}, \cdots, p_{n,F}]=diag(\bs{s}_n^H\bs{s}_n)$, where $\bs{p}_n\in  \mathbb{R}^F$. Note that $H$ is the Hermitian transpose, and then $p_{n,f}= |s_{n,f}|^2 = {a_{n,f}}^2 + {b_{n,f}}^2 \in \mathbb{R}$. Finally, cepstral feature $\bs{c}_n\in \mathbb{R}^M$, where $M<F$, is acquired after inverse DFT plus filtering and frequency warping on log-amplitude spectrum $[\log \sqrt{p_{n,1}}, \cdots, \log \sqrt{p_{n,F}}]$. Notice that $\bs{p}_n$ encodes the amplitude of $\bs{s}_n$ while ignoring the phase. 
Therefore, $\bs{s}_n$ and the waveform cannot be accurately reproduced only given $\bs{c}_n$. 

Conventional vocoders assume that $\bs{s}_{n}$ has a minimum phase in order to approximately recover $\bs{s}_n$ from $\bs{c}_n$. However, the speech waveform is not merely a minimum-phase signal. One solution is to model the phase as additional acoustic features \cite{6288938,Degottex2014}. Another way is to directly model complex-valued $\bs{s}_n$ by using complex-valued NN \cite{7472755} or restricted Boltzman machine \cite{Nakashika2017}.
Alternatively, a statistical model such as the Wavenet-based vocoder \cite{Tamamori2017} can be used rather than using a deterministic vocoder. Such a vocoder may learn the phase information not encoded by the acoustic features. 

This work used a conventional vocoder called WORLD \cite{morise2016world} as the baseline. It then included a phase-recovery technique \cite{griffin1984signal}, a waveform synthesizer based on a log-domain pulse model \cite{degottex2017log}, and a Wavenet-based vocoder for comparison. Complex-valued approaches may be included in future work. 
}

\subsubsection{Deterministic vocoders}

{
The WORLD vocoder \cite{morise2016world} is similar to the legacy STRAIGHT vocoder \cite{ref:Kawahara99} and uses a source-filter-based speech production model with mixed excitation. It also assumes a minimum phase for the spectrum. By using the procedure in Sec.~\ref{cep-conversion} reversely, WORLD converts the cepstral features, $\bs{c}_n$, given by the acoustic model into a linear amplitude spectrum. 
Then, WORLD creates the excitation signal by mixing a pulse and a noise signal in the frequency domain, where each frequency band is weighted by a BAP value. Finally, it generates a speech waveform based on the source-filter model. 

Since WORLD assumes a minimum phase, this work included a phase-recovery method that may enhance the phase of the generated waveform by WORLD. This method re-computes the amplitude spectrum given the generated waveform and then applies the Griffin-Lim phase-recovery algorithm. Note that this method is different from that in our previous work \cite{Takaki2017,Kaneko2017} where phase-recovery was directly conducted on the generated amplitude spectrum. 

This work also included the pulse model in the log-domain (PML) \cite{degottex2017log} as another waveform generator. One advantage of PML is that it avoids voicing decision per frame and may alleviate the perceptual degradation caused by voicing decision errors. In addition, PML may better represent noise in voiced segments. The PML and WORLD in this work used the same generated spectrum as input. The noise mask used by PML is generated by another RNN given input linguistic features, which it is not indicated in Fig.~\ref{fig:systems}.
}

\subsubsection{Data-driven vocoder}
\label{sec:wavent}
Besides the deterministic vocoders, this work included a statistical vocoder based on Wavenet \cite{Tamamori2017, ustcbc2017}. 
Suppose $\bs{o}_{t}$ is a one-hot vector representing the quantized waveform sampling point at time $t$, the Wavenet vocoder infers the distribution $P(\bs{o}_{t}|\bs{o}_{t-R:t-1}, \bs{a}_t)$ given the acoustic feature $\bs{a}_t$ and previous observations $\bs{o}_{t-R:t-1}=\{\bs{o}_{t-R}, \cdots, \bs{o}_{t-1}\}$, where $R$ is the receptive field of the network.  
In this work, ${\bs{a}}_t$ contained the F0 and MGC.
Note that, while natural $\bs{a}_t$ is used to train the vocoder, $\widehat{\bs{a}}_t$ generated by the acoustic model is used for waveform generation.

\begin{figure*}[t]
\includegraphics[width=\textwidth]{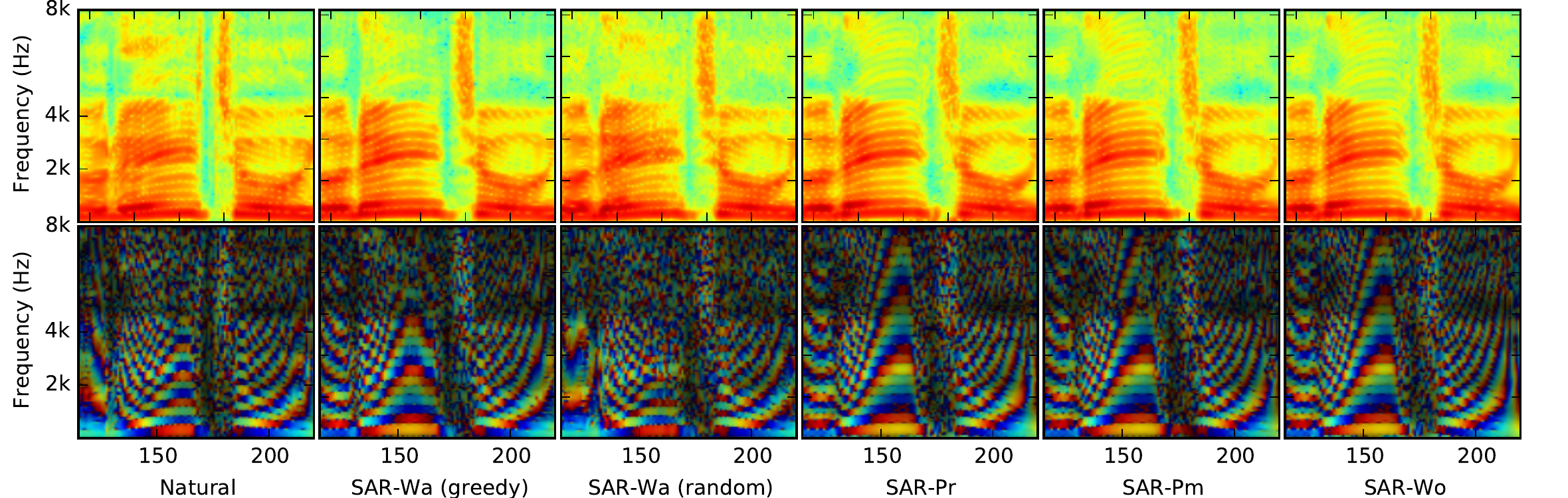}
\vspace{-6mm}
\caption{Spectrogram (top) and instantaneous frequency (bottom) of natural speech sample, synthetic samples from \texttt{SAR-Wa} using greedy generation, \texttt{SAR-Wa} using random sampling, \texttt{SAR-Pr}, \texttt{SAR-Pm} and \texttt{SAR-Wo}. Test utterance ID is AOZORAR\_03372\_T01.}
\vspace{-2mm}
\label{fig:genMethod}
\end{figure*}

During generation, $\widehat{\bs{o}}_{t}$ is usually obtained by random sampling, i.e., $\widehat{\bs{o}}_{t} \sim P(\bs{o}_{t}|\widehat{\bs{o}}_{t-R:t-1}, \widehat{\bs{a}}_t)$. However, we found that the synthetic waveform sounded better when samples in the voiced segments were generated as $\widehat{\bs{o}}_{t} = \arg\max_{\bs{o}_{t}}p(\bs{o}_{t}|\widehat{\bs{o}}_{t-R:t-1}, \widehat{\bs{a}}_t)$. This method is referred to as the \emph{greedy generation} method. 
Fig.~\ref{fig:genMethod} plots the spectrogram and the instantaneous frequency (IF) \cite{boashash1992estimating,engel17a} of generated waveforms from the two methods, i.e., \texttt{SAR-Wa} (greedy) and \texttt{SAR-Wa} (random). Compared with random sampling, the greedy method generated more regular IF patterns in voiced segments. It perceptually made the waveforms sound less trembling. 
One reason for this may be that the `best' generated $\widehat{\bs{o}}_{t}$ was temporally more consistent with $\widehat{\bs{o}}_{t-R:t-1}$. 
Note that samples in the unvoiced parts were still randomly drawn, and the voiced/unvoiced state for each time could be inferred from the F0 in $\widehat{\bs{a}}_t$.

\subsection{Neural network acoustic models}
\label{sec:acous}
The NN-based acoustic models included in this work aims at inferring the distribution of the acoustic feature sequence $\bs{a}_{1:N}=\{\bs{a}_1, \cdots, \bs{a}_N\}$ conditioned on the linguistic feature $\bs{l}_{1:N}=\{\bs{l}_1, \cdots, \bs{l}_N\}$ in $N$ frames. 
The baseline RNN, whether it has a recurrent output layer or not, defines the distribution as
\vspace{-2mm}
\begin{equation}
p(\bs{a}_{1:N}|\bs{{l}}_{1:N}) ={\prod_{n=1}^{N}p(\bs{a}_{n}|\bs{l}_{1:N})}=\prod_{n=1}^{N}\mathcal{N}(\bs{a}_n; \bs{h}_n, \bs{I})
\label{eq:RNN}
\vspace{-2mm},
\end{equation}
where $\mathcal{N}(\cdot)$ is the Gaussian distribution, $\bs{I}$ is the identity matrix, $\bs{h}_n=\mathcal{H}_{\bs{\Theta}}(\bs{l}_{1:N}, n)$ is the outcome of the output layer at frame $n$, and $\bs{\Theta}$ is the network's weight. 
For generation, $\widehat{\bs{a}}_{1:N}$ can be acquired by using a mean-based method that sets $\widehat{\bs{a}}_{1:N} =\bs{h}_{1:N}$. 

A shortcoming of RNN is that the across-frame dependency in $\bs{a}_{1:N}$ is ignored. Hence, this work included a model that takes into account the dependency in a causal direction  \cite{wangARRMDN}. 
This model, which is called shallow AR ({SAR}), defines the distribution as:
\vspace{-1mm}
\begin{align}
p(\bs{a}_{1:N}|\bs{{l}}_{1:N}) &= {\prod_{n=1}^{N}p(\bs{a}_{n}|\bs{a}_{n-K:n-1}, \bs{l}_{1:N})} \\
&=\prod_{n=1}^{N}\mathcal{N}(\bs{a}_n; \bs{h}_n + \mathcal{F}_{\Phi}(\bs{a}_{n-K:n-1}), \bs{I}).
\vspace{-6mm}
\label{eq:SAR}
\end{align}
This model uses $\mathcal{F}_{\bs{\Phi}}(\bs{a}_{n-K:n-1}) = \sum_{k=1}^{K}\bs{\beta}_k\odot\bs{a}_{n-k}+\bs{\gamma}$ to merge the acoustic features in the previous $K$ frames and then changes the distribution for frame $n$, which builds the causal across-frame dependency. Another deep AR (DAR) model similar to SAR can be defined \cite{wangDARF0}. DAR in this work was only used to model the quantized F0 for all the experimental methods, and thus is not explained here.

As RNN and SAR are trained by using the maximum-likelihood criterion and then generate by using the mean-based method, they may not produce acoustic features with natural textures.
Therefore, this work also included the GAN-based postfilter \cite{kaneko2017generative} to enhance RNN and SAR. 
The GAN discriminators in this work did not crop the input acoustic features into small patches, which is different from original work. Instead, they made a true-false judgment every frame.

\section{Experiments}
\label{sec:exp}
\subsection{Corpus and features}
This work used a Japanese speech corpus \cite{kawai2004ximera} of neutral reading speech uttered by a female speaker. The duration is 50 hours, and the number of utterance is 30,016, out of which 500 were randomly selected as a validation set and another 500 were used as a test set.
Linguistic features were extracted by using OpenJTalk \cite{HTSWorkingGroup2014JOPEN}. The dimension of these features vectors was 389. Acoustic features were extracted at a frame rate of 200 Hz (5 ms) from the 48 kHz waveforms. MGC and BAP were extracted by using WORLD. The dimensions for MGC were 60 and those for BAP were 25. The F0 was extracted by an assembly of multiple pitch trackers and then quantized into 255 levels for quantized F0 modeling \cite{wangDARF0}.

\begin{figure}[t]
\includegraphics[width=\columnwidth]{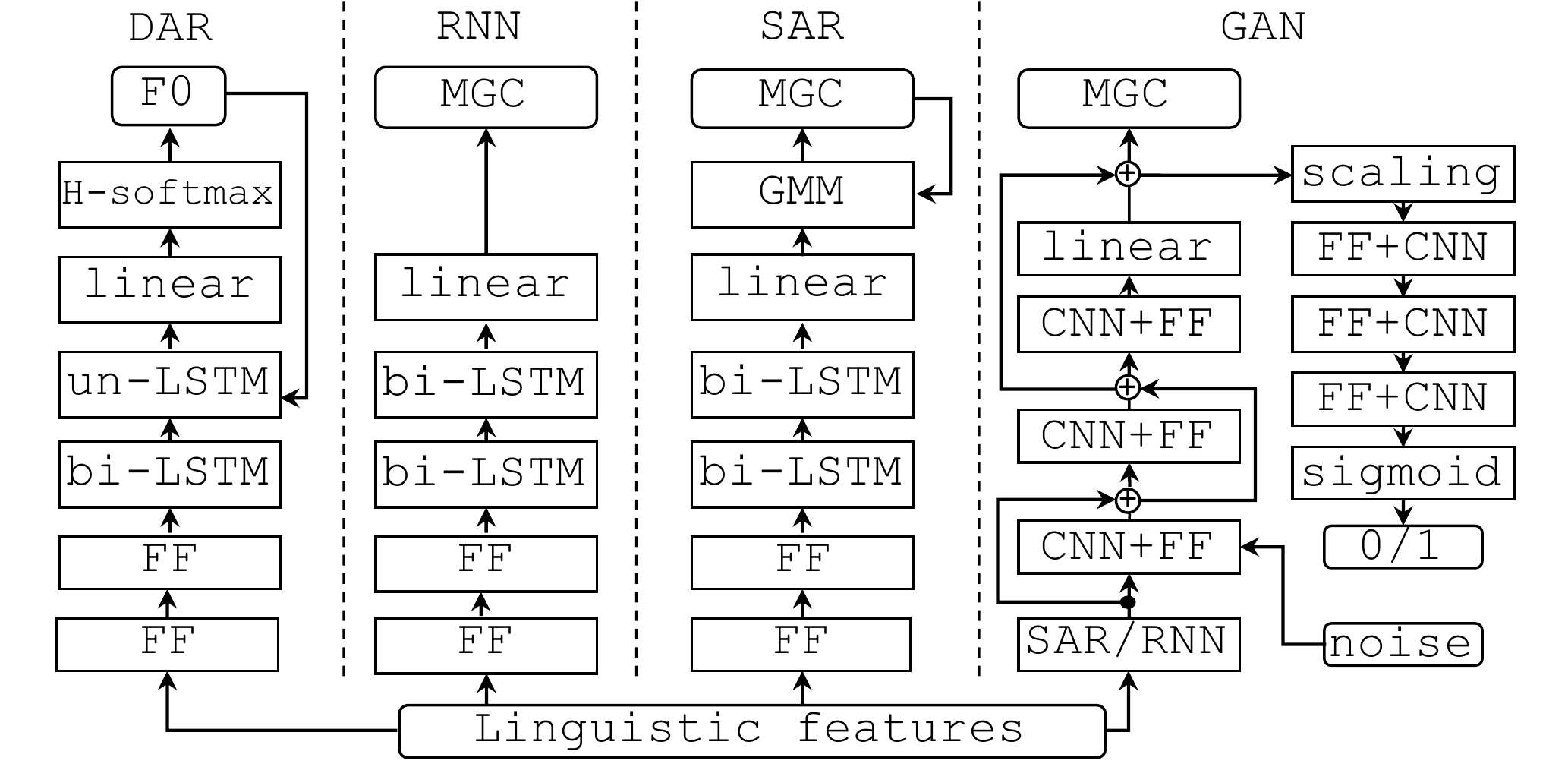}
\vspace{-2mm}
\caption{Structure of acoustic models. 
Linear, FF, H-softmax, and un- and bi-LSTM correspond to linear transformation, feedforward layer using $tanh$, hierarchical softmax \cite{wangDARF0}, and uni- and bi-directional LSTM layers. GMM is Gaussian mixture model. Scaling layer in {GAN} scales each dimension of input features.}
\label{fig:network}
\end{figure}

\subsection{Neural network configuration}
The network structures of the acoustic models are plotted in Fig.~\ref{fig:network}. In {DAR}, {RNN} and {SAR}, the size of feedforward (FF),  bi- and uni-directional LSTM layer was 512, 256, and 128. The size of the linear layer depended on the output features' dimensions. For {SAR}, the output layer included a Gaussian distribution for MGC with the AR parameter $K=1$ and another Gaussian for BAP with $K=0$. 
In {GAN}, all the CNN layers had 256 output channels and conducted 1D convolution. 
Each FF layer in the GAN generator changed the dimension of a CNN's output before skip-add operation. 
To reduce instability in low-dimensional MGCs, the input MGC to the discriminator was scaled element-wisely. The scaling weight was $0.001$ for the first five MGC dimensions, $0.01$ for the next five dimensions, and $1$ for the rest. The weight was $0$ for BAP.

The Wavenet vocoder works at a sampling rate of 16kHz. The $\mu$-law companded waveform is quantized into 10 bits per sample. 
Similar to the literature \cite{Tamamori2017}, the network consists of a linear projection input layer, 40 blocks for dilated convolution, and a post-processing block. The $k$-th dilation block has a dilation size of $2^{\mathrm{mod}(k,10)}$, where $\mathrm{mod}(\cdot)$ is  modulo operation. The acoustic features, which are fed to every dilated convolution block, contains MGC and quantized F0. Natural acoustic features are used for training while generated MGC and F0 are used during generation. 

All the acoustic models and Wavenet are implemented on a modified CURRENNT toolkit \cite{weninger2015introducing}. This toolkit and training recipes can be found online (\url{http://tonywangx.github.io}). 

\begin{figure}[t]
\includegraphics[width=\columnwidth]{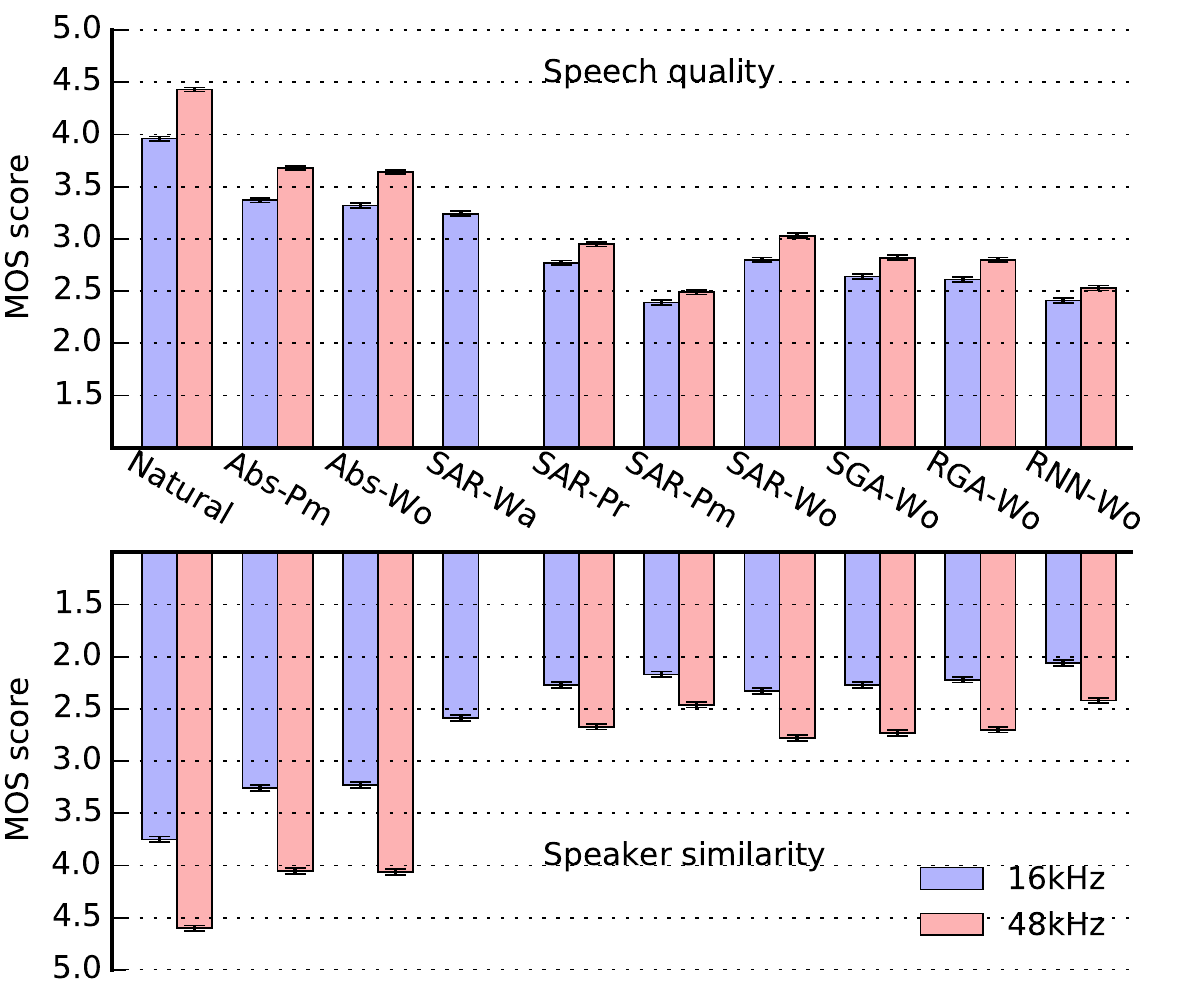}
\vspace{-6mm}
\caption{Results in MOS scale. \texttt{AbS-Pm} and \texttt{AbS-Wo} correspond to vocoded speech by using PML and WORLD vocoders. Error bars represent two-tailed Student 95\% confidence intervals.}
\vspace{-2mm}
\label{fig:mos}
\end{figure}

\subsection{Evaluation environment}
The seven speech synthesis methods in Fig.\ref{fig:systems} were evaluated in terms of speech quality and speaker similarity. 
Natural and vocoded speech from WORLD and PML were also included in the evaluation.
All the systems except for the Wavenet-based \texttt{SAR-Wa} were rated at sampling rates of 48 and 16 kHz.
The speech samples of these systems were natural or generated at 48 kHz, and the 16 kHz samples were down-sampled from these 48 kHz samples.
\texttt{SAR-WA} only generated samples of 16 kHz. All samples were normalized to -26 dBov, and a total of 19 groups of speech samples were evaluated. 

The evaluation was crowdsourced online. Each evaluation set had 19 screens, i.e., one for each system. The order of systems in each set was randomly shuffled. The evaluators must answer two questions on each screen. First, they listened to a sample of the system under evaluation and rated the naturalness on a 1-to-5 MOS scale. They then rated the similarity of that test sample to the natural 48 kHz sample on a 1-to-5 MOS scale. The participants were allowed to replay the samples. 
The samples in one set were synthesized for the same text randomly selected from the test set.

\subsection{Results and discussion}
\label{sec:dis}

\begin{figure}[t]
\includegraphics[width=\columnwidth]{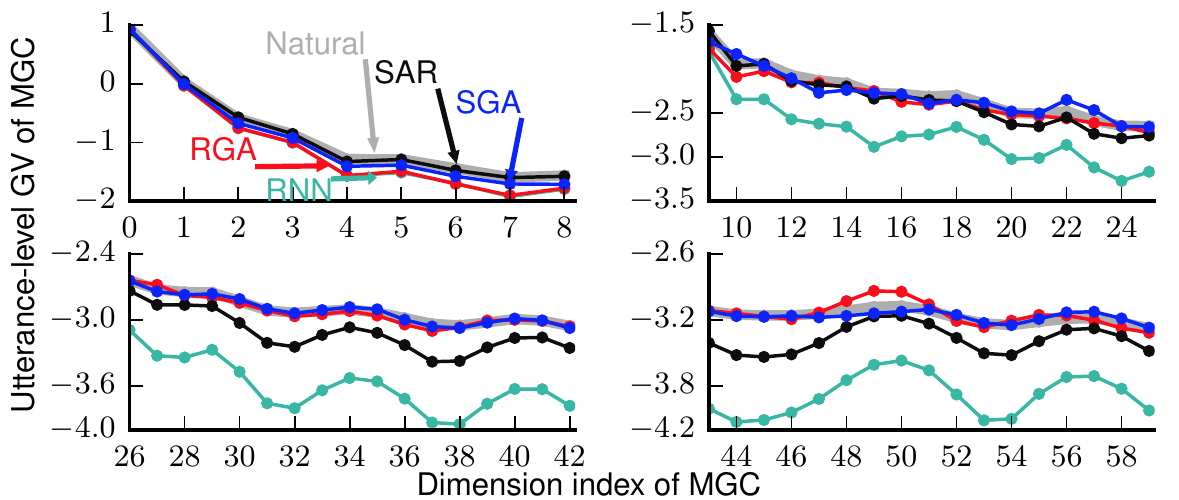}
\vspace{-4mm}
\caption{Global variance (GV) of natural and generated MGC averaged on test set. }
\vspace{-2mm}
\label{fig:gv}
\end{figure}

\begin{figure}[t]
\includegraphics[width=\columnwidth]{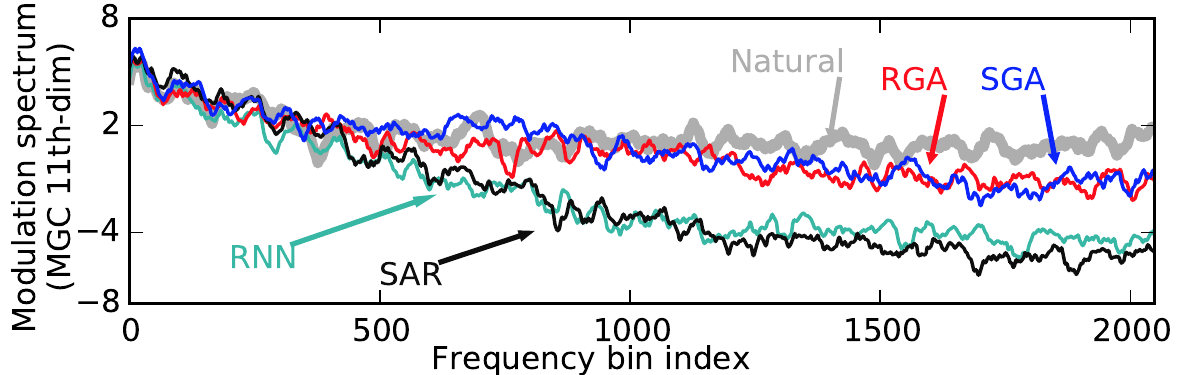}
\vspace{-5mm}
\caption{Modulation spectrum of 11th dimension of MGC for test utterance AOZORAR\_03372\_T01.}
\vspace{-2mm}
\label{fig:ms}
\end{figure}

We collected a total of 1500 evaluation sets, i.e., 1500 scores for each system. 
A total of 235 native Japanese listeners participated, with an average of 6.4 sets per person. 
The statistical analysis was based on unpaired t-tests with a 95\% confidence margin and Holm-Bonferroni compensation.  
The results are plotted in Fig.~\ref{fig:mos}.

{
On acoustic modeling, the comparisons of the speech quality score among \texttt{RNN-Wo}, \texttt{RGA-Wo}, \texttt{SAR-Wo}, and \texttt{SGA-Wo} indicated that {SAR} (3.03) $>$ {SGA} (2.82) $\sim$ {RGA} (2.81) $>$ {RNN} (2.53) for both 48 kHz and 16 kHz. The same ordering can be seen in terms of speaker similarity.
RNN's unsatisfactory performance may be due to the over-smoothing effect on the generated MGC, which muffled the synthetic speech. This effect is indicated by the global variance (GV) plotted in Fig.~\ref{fig:gv}, where the GV of the generated MGC from RNN was smaller than that of the other models. 
The performance of SAR is consistent with that in our previous work \cite{wangARRMDN}, which indicated that the AR model alleviated the over-smoothing effect.
The GAN-based postfilter in both SGA and RGA also reduced the impact of over-smoothing. 
Further, GAN not only enhanced GV but also compensated for the modulation spectrum (MS) of the generated MGC throughout the whole frequency band, which can be seen from Fig.~\ref{fig:ms}.
However, neither SGA nor RGA outperformed SAR even though SAR did not boost the MS in the high frequency bands.
After listening to the samples, we found that, while the samples of SGA and RGA had better spectrum details, they contained more artifacts. This indicated that generators in SGA and RGA may not optimally learn the distribution of natural MGC. Future work will look into details on SGA and RGA.

On waveform generation methods, the comparison among \texttt{SAR-Wa}, \texttt{SAR-Pr}, \texttt{SAR-Pm}, \texttt{SAR-Wo}, and vocoded speech showed that \texttt{SAR-Wa} significantly outperformed other waveform generation methods in terms of speech quality. More interestingly, \texttt{SAR-Wa}'s quality was even judged to be  better than other synthetic methods by using a 48 kHz sampling rate. The quality of \texttt{SAR-Wa} was also close to the 16 kHz vocoded samples from \texttt{Abs-Pm} and \texttt{Abs-Wo}. 
A closes inspection of the samples showed that the Wavenet vocoder in \texttt{SAR-Wa} had fewer artifacts, such as amplitude modulation in the waveform and other effects possibly due to conventional vocoders. 
However, the gap between \texttt{SAR-Wa} and \texttt{Abs-Wo}/\texttt{Abs-Pm} was still large in terms of speaker similarity. 
One reason may be that, while the Wavenet vocoder trained by using natural acoustic features may well learn the mapping from acoustic features to waveforms, during waveform generation it cannot compensate for the degradation of generated acoustic features caused by the imperfect acoustic model \texttt{SAR}. 
Note that this problem may not be resolved by training the Wavenet vocoder using generated acoustic features if the acoustic model is not good enough. 
 

Among other waveform generation methods, the PML vocoder, while being similar to WORLD for analysis-by-synthesis, it lagged behind when using the generated acoustic features. This result is somewhat different from that in \cite{degottex2017log}, and future work will investigate the reasons for this.
Comparison between \texttt{SAR-Pr} and \texttt{SAR-Wo} indicated that the phase recovery method did not improve the quality or similarity score. 
We suspect that this is because the iterative process in the Griffin-Lim algorithm may introduce some artifacts that degraded speech quality. 
}

\section{Conclusion}
\label{sec:con}
This work was our initial step in building a framework in which recent acoustic modeling and waveform generation methods could be compared on a common ground by using a large-scale perceptual evaluation. This work only considered a few methods that could easily be plugged into the common TTS pipeline. 
On acoustic models, the results showed that the autoregressive model {SAR} could achieve a better performance than a normal RNN. This SAR was also easier to train than a GAN-based model. On waveform generation methods, this work demonstrated the potential of the Wavenet vocoder and the advantage of statistical waveform modeling compared with conventional deterministic approaches.

We intend to investigate the reasons for the experiments' results in more detail in future work. 
Meanwhile, we recently validated that \texttt{SAR} is a simple case of using normalizing flow \cite{rezende2015variational} to transform the density of target data. We will try to improve \texttt{SAR} and the overall performance of speech synthesis system.
Complex-valued models that support the modeling of waveform phases may also be included.


\vfill\pagebreak

\bibliographystyle{IEEEbib}
\bibliography{bib}

\end{document}